\documentclass[modern]{aastex701}
\usepackage{graphicx}
\usepackage[margin=1in, letterpaper]{geometry}
\usepackage{amsmath}
\usepackage{booktabs}

\newcommand{\logg}{\log(g)}
\newcommand{\teff}{T_{\rm eff}}
\newcommand{\DAge}{\Delta \text{Age}}
\newcommand{\DLogAge}{\Delta \log_{10}(\text{Age})}
\newcommand{\vsini}{v\sin(i)}

\newcommand{\Rguide}{R_{\rm guide}}
\newcommand{\Zmax}{Z_{\rm max}}

\newcommand{\KSabdMAC}{\rm{KS}_{\rm abd}^{\rm MAC}}
\newcommand{\KSabdMDC}{\rm{KS}_{\rm abd}^{\rm MDC}}
\newcommand{\KSageMAC}{\rm{KS}_{\rm age}^{\rm MAC}}
\newcommand{\KSageMDC}{\rm{KS}_{\rm age}^{\rm MDC}}
\newcommand{\logp}{\log_{10}(p)}

\submitjournal{ApJ}

\sloppypar

\shorttitle{evolved stars with inconsistent age estimates}
\shortauthors{frazer, griffith, hogg}
\begin{document}
\title{Evolved stars with inconsistent age estimates:\\ Abundance outliers or mass transfer products?}

\author[0009-0003-3109-1216]{Polly Frazer}
\affiliation{Center for Cosmology and Particle Physics, Department of Physics, New York University, 726~Broadway, New~York,~NY 10003, USA}
\email{pollyfrazer@nyu.edu}

\author[0000-0001-9345-9977]{Emily J. Griffith}
\altaffiliation{Hubble Fellow}
\affiliation{Center for Astrophysics and Space Astronomy, Department of Astrophysical and Planetary Sciences, University  of Colorado, 389~UCB, Boulder,~CO 80309-0389, USA}
\email{Emily.Griffith-1@colorado.edu}

\author[0000-0003-2866-9403]{David W. Hogg}
\affiliation{Center for Cosmology and Particle Physics, Department of Physics, New York University, 726~Broadway, New~York,~NY 10003, USA}
\affiliation{Max-Planck-Institut f{\"u}r Astronomie, K{\"o}nigstuhl 17, D-69117 Heidelberg, Germany}
\affiliation{Center for Computational Astrophysics, Flatiron Institute, 162 Fifth Avenue, New York, NY 10010, USA}
\email{david.hogg@nyu.edu}

\author[0009-0005-0182-7186]{Amaya Sinha}
\affiliation{Department of Physics \& Astronomy,  University of Utah, 270 South 1400 East, Salt Lake City, UT 84112, USA}
\email{u1363702@utah.edu}

\author[0000-0002-4818-7885]{Jamie Tayar}
\affiliation{Department of Astronomy, University of Florida, Gainesville, FL 32611, USA}
\email{jtayar@ufl.edu}

\begin{abstract}\noindent
In the Milky Way disk there is a strong trend linking stellar age to surface element abundances.
Here we explore this relationship with a dataset of 8,803 red-giant and red-clump stars with both asteroseismic data from NASA \textsl{Kepler} Mission and surface abundances from the \textsl{SDSS-V MWM}.
We find, with a k-nearest-neighbors approach, that the [Mg/H] and [Fe/Mg] abundance ratios predict asteroseismic ages to an accuracy of about 2 Gyr for the majority of stars.
That said, there are substantial outlier stars whose surface abundances do not match their asteroseismic ages.
Because asteroseismic ages for these stars are fundamentally based on density or mass, these outliers are mass-transfer candidates.
Stars whose surface abundances predict a younger age (higher mass) than what's seen in the asteroseismology are \emph{mass accretor candidates} (MAC);
stars whose abundances predict an older age (lower mass) than the asteroseismology age are \emph{mass donor candidates} (MDC).
We create precise control samples, matched according to (1) surface abundances and (2) asteroseismic ages, for both the MAC and MDC stars;
we use these to find slight differences in rotational velocity, [C/N], and [Na/Mg] between the mass-transfer candidates and their abundance neighbors.
We find no drastic differences in kinematics, orbital invariants, UV excess, or other stellar abundances between outliers and their abundance neighbors.
We deliver 377 mass-transfer candidates for follow-up observations.
This project implicitly suggests a fundamental limit on the reliability of asteroseismic ages, and supports existing evidence that age-abundance outliers are products of binary mass transfer. 
\end{abstract}

\section{Introduction}

While stellar ages can not be directly measured, many methods exist to infer age from correlated stellar parameters, model fitting, and empirical observational relationships. Age indicators span spectroscopic abundance diagnostics, including lithium depletion \citep{burke2004, mentuch2008, soderblom2010}; carbon to nitrogen ratios \citep{masseron2015, ness2016, martig2016}; and metallicity \citep{wheeler1989, feltzing2009}, and photometric relationships, such as isochrone fits to stellar temperature and luminosity \citep{edvardsson1993, ng1998, pont2004, jorgensen2005, serenelli2013} and model fits to asteroseismic oscillation modes in stars \citep{miglio2013, pinsonneault2014}. However, no age diagnostic is universal. Instead, each method can infer ages for a subset of stellar states, masses, and/or ages while also requiring a specific type, quality, and/or quantity of observational data \citep[e.g.,][]{soderblom2010}.

The [$\alpha$/Fe] and [Fe/H] abundances, for example, are a relative age indicator for many main sequence and evolved stars that can easily be measured with low to high-quality spectral data \citep[e.g.,][]{wyse1988, bovy2012, feltzing2009}. At early times, stars are enriched primarily by massive stars, which quickly exhaust their nuclear fuel and eject large amounts of $\alpha$ elements (like Mg and O) and some Fe upon their explosive deaths as core-collapse supernovae (CCSN). On more extended timescales, Type-Ia supernovae (SNIa)---the thermonuclear explosions of white dwarf binary systems---produce large amounts of Fe, increasing [Fe/H] and decreasing the [$\alpha$/Fe] ratio with time \citep[e.g.,][]{andrews2017}. Consequently, older stars will tend to have enhanced [$\alpha$/Fe] (or depleted [Fe/$\alpha$]) and lower metallicity as compared to younger stars. While not the most precise age indicator ($\sim2-3$ Gyr precision), this abundance age can be inferred for millions of stars with spectroscopic data, and is often leveraged in data-driven age diagnostics \citep{ness2016, xiang2017, 2023MNRAS.522.4577L, wang2024}. 

Conversely, asteroseismic ages, inferred from density estimates derived from the oscillation modes of stars \citep{brown1991, kjeldsen1995}, are very precise (within $\sim$10\%) but require extensive observational resources \citep{soderblom2010}. Large, space-based, time-domain surveys including CoRoT \citep{baglin2003}, \textit{Kepler} \citep{borucki2010, Yu_2018}, K2 \citep{howell2014}, and TESS \citep{ricker2015} have enabled the measurement of these oscillation modes in $\sim$tens of thousands of solar-like and evolved stars over the last two decades. Coupled with spectroscopic surveys such as the Apache Point Observatory Galactic Evolution Experiment \citep[APOGEE;][]{majewski2017} and the GALactic Archaeology with HERMES survey \citep[GALAH;][]{desliva2015, buder2021, buder2024} asteroseismic parameter can be combined with spectroscopic measurements to infer masses, radii, and ages for large stellar populations \citep{miglio2013, pinsonneault2014, mackereth2021, zinn2022, warfield2024, pinsonneault2024apokasc3jointspectroscopicasteroseismic}.

With large-scale space and ground-based surveys, the sample of stars for which we have age estimates from multiple methods is growing. We find, however, that ages derived from different methods do not always agree \citep[e.g.,][]{soderblom2010, stonemartinez2025}. In addition to smaller, systematic differences in age diagnostics, there exist smaller samples of stars whose abundance ages and asteroseismic ages drastically differ. These populations include the young $\alpha$-rich (YAR) stars, identified by \citet{2015MNRAS.451.2230M} and \citet{Chiappini2015}, and a small population of very old $\alpha$-poor stars, some of which have astroseismic ages older than the age of the universe  \citep[e.g.,][]{2023A&A...671A..21J,pinsonneault2024apokasc3jointspectroscopicasteroseismic}. 

YAR objects have high [$\alpha$/Fe] abundances, typical of old stars, but young astroseismic ages. Such objects are not predicted to exist by most Galactic Chemical Evolution (GCE) models \citep[e.g.,][]{haywood2016}, though some models \citep[e.g.,][]{johnson2021, sun2023} find evidence of $\alpha$-enhanced stars that are truly young. Many follow-up observational studies have investigated the origin of these young, massive stars---studying their kinematics \citep{2016A&A...595A..60J, 2023A&A...671A..21J, silva2018, 2021ApJ...922..145Z}, chemistry \citep{2019Hekker, 2021ApJ...922..145Z, grisoni2024}, and magnetic activity \citep{yu2024newevidencebinarityyoung}. While some stars appear to be truly young \citep[e.g.,][]{lu2025}, many show signs of binary, such as radial velocity scatter, and hint at ongoing or past binary mass transfer \citep{2023A&A...671A..21J}. As asteroseismic ages are derived from stellar masses and densities, an inflated mass from binary mass transfer would result in an inferred asteroseismic age that is artificially young.
Conversely, the outlying old $\alpha$-poor stars have near-solar [$\alpha$/Fe] abundances, typical of young stars, but old astroseismic ages. These stars have not been followed up as heavily as the YAR population, but are equally interesting outliers that likely arise due to mass loss from binary interaction \citep{li2022, 2023A&A...671A..21J, pinsonneault2024apokasc3jointspectroscopicasteroseismic}.

To date, the studies of such stars with discrepant abundance and asteroseismic ages have been restrictive. Follow-up long-term radial velocity (RV) monitoring campaigns \citep{2016A&A...595A..60J, 2023A&A...671A..21J} and detailed analyses of C, N, and O abundances \citep{2019Hekker} focus on small samples ($< 50$) of YAR stars. The YAR targets are typically selected with strict abundance and age thresholds, limiting follow-up to the most massive stars ($M > 1.3 M_{\odot}$) in the high-$\alpha$ population. While these works have yielded interesting results for individual cases, a statistical analysis of stars with discrepant asteroseismic and abundance ages is needed.

In this paper, we systematically identify stars whose abundance and asteroseismic ages are in disagreement. By searching the full abundance-age plane, our selection is not limited to only the young asteroseismic (most massive) $\alpha$-rich stars. As past studies have provided some evidence that these age outliers are the result of binary interactions that strip or accrete mass, we search for statistical differences in the distributions of stellar parameters between the sample of age-discrepant stars and a control population---focusing on stellar parameters that may be indicative of past or ongoing binary interactions. In Section~\ref{sec:data}, we describe the age and abundance catalogs used in this study. We infer ages from the stellar [Fe/Mg] and [Mg/H] abundances in Section~\ref{sec:abund_age} and identify stars with discrepant abundance and asteroseismic ages in Section~\ref{outlier_section}. In Section~\ref{sec:explor_params}, we construct a control sample and compare distributions of astrometric parameters, rotational velocities, elemental abundances, UV fluxes, and orbital invariants between the age-discrepant stars and two control samples. Finally, we discuss our results in Section~\ref{sec:discussion} and highlighting a few interesting objects.

\section{Data Samples}\label{sec:data}

In this work, we leverage asteroseismic and spectroscopic stellar parameters for stars in the APOKASC-3 catalog \citep[][hereafter P24]{pinsonneault2024apokasc3jointspectroscopicasteroseismic}.
This catalog combines spectroscopic data from the 16th and 17th data releases of APOGEE-2 (\citealp{ahumada2020} DR16; \citealp{Abdurro_uf_2022} DR17), part of the fourth generation of the Sloan Digital Sky Survey \citep[SDSS-IV;][]{blanton2017, majewski2017} with asteroseismic data from the \textit{Kepler} \citep{borucki2010, Yu_2018}. 
\textit{Kepler}, NASA’s first exoplanet search mission, observed stellar brightness variations over extended periods, capturing long-cadence oscillations in stars. Regular patterns of radial modes can be identified in the oscillation data for giant stars, yielding measurements of $\Delta\nu$ (frequency spacing) and $\nu_{max}$ (frequency of maximum power), from which stellar densities can be derived \citep[e.g.,][P24]{Yu_2018}.

The complementary APOGEE-2 \citep{majewski2017} survey has taken high-resolution (R $\sim 22,500$), near-infrared spectra of hundreds of thousands of stars with the twin spectrographs on the 2.5m Sloan Foundation telescope \citep{gunn2006, wilson2019} at Apache Point Observatory and the 2.5m du Pont Telescope \citep{bowen1973} at Las Campanas Observatory. Using the APOGEE data processing pipeline \citep{nidever2015} and the APOGEE Stellar Parameter and Chemical Abundance Pipeline \citep[ASPCAP;][]{holtzman2015, garcia_perez2016}, APOGEE DR16 and DR17 provide detailed measurements of stellar atmospheric parameters (including $\teff$ and $\logg$) and abundances (including [Fe/H], [Mg/Fe], [C/Fe], and [N/H]. See \citet{jonsson2020} for additional details on the APOGEE spectroscopic analysis.

Combining stellar parameters measured by APOGEE with the asteroseismic measurements from \textit{Kepler}, APOKASC-3 derives stellar densities, masses, radii, and ages. Specifically, mean stellar densities are inferred from measurements of $\Delta \nu$ and $\nu_{\rm max}$ from \textit{Kepler} and spectroscopic $\teff$ from APOGEE. Stellar radii, surface gravities, and masses are then computed from scaling relations, and ages are inferred from stellar models using Bayesian inference (P24). APOKASC-3 reports asteroseismic parameters for 12,448 stars. For a detailed discussion of the \textit{Kepler}-APOGEE targeting, see \citet{simonian2019} and P24.

For our analysis, we employ the APOKASC-3 stellar ages alongside stellar parameters from the Milky Way Mapper (MWM) survey---the next generation stellar spectroscopic survey within SDSS-V \citep{kollmeier2017, kollmeier2025}. In contrast to previous iterations of SDSS, SDSS-V operates with a new robotic Focal Plane System \citep{pogge2020, sayres2022}, upgraded data reduction pipeline (Nidever et al., in prep), and new analysis framework (Astra; Casey et al., in prep), which implements ASPCAP and other algorithms. We use data from the 19th data release of SDSS (DR19), the first major release of MWM data products \citep{meszaros2025, sdssdr19}\footnote{Specifically, we use the \texttt{astraAllStarASPCAP-0.6.0} data.}. While the APOKASC-3 catalog utilizes spectroscopic values from previous APOGEE Data Releases, this work leverages the updated stellar parameters and stellar abundances derived from the upgraded SDSS-V DR19 analysis. Our results do not change if we repeat our analysis with SDSS-IV DR17 products.

To select the stars with the highest quality asteroseimic and spectroscopic data, we apply quality cuts based on the APOKASC-3 and MWM data flags.
In APOKASC-3, the data are classified into three quality categories for age determinations: Gold, Silver, and Detection. Gold stars have the highest precision and accuracy, with both $\Delta\nu$ and $\nu_{max}$ measured. Silver stars have less precision, with larger age uncertainties, but still have measurements for both $\Delta\nu$ and $\nu_{max}$. Detection stars have only $\nu_{max}$ measured, leading to a lower age accuracy. We select only stars in the Gold category ($N=10,036$), as their age estimates are the most trustworthy, and further restrict our sample to stars with age errors less than 2 Gyrs.
When the evolutionary state of the star is uncertain and multiple stellar ages are provided, we adopt the RGB age and associated uncertainties. 

\begin{figure}
    \centering
    \includegraphics[width=\linewidth]{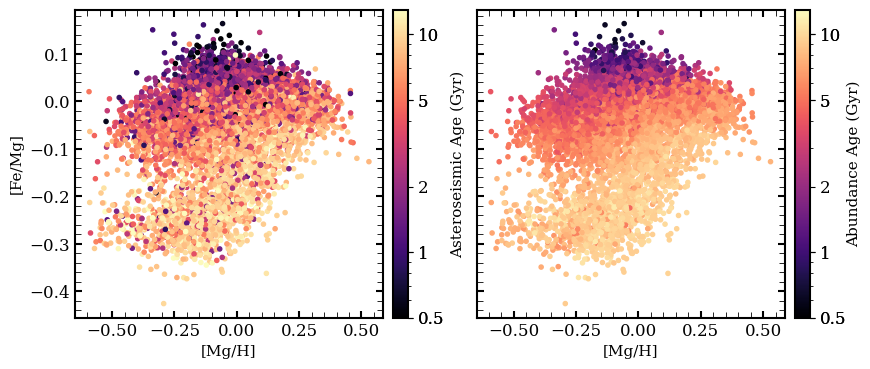}
    \caption{Left: [Mg/H] vs. [Fe/Mg] for our sample, colored by asteroseismic age. Younger stars (darker colored) tend to have higher [Fe/Mg], while older stars (lighter colored) tend to have lower [Fe/Mg]. Right: Same as left panel, but now colored by our abundance ages.}
    \label{fig:mg_vs_fe_log_age}
\end{figure}

We apply additional quality cuts based on the MWM catalog by removing stars with \texttt{mg\_h\_flag} and \texttt{fe\_h\_flag} set, indicative of unreliable [Fe/H] or [Mg/H]. We remove duplicate observations, keeping the observation with the highest signal to noise. 
Finally, we restrict our sample to stars with [Mg/H] $> -0.6$ dex.
Our final sample consists of 8,803 stars with high-quality ages from APOKASC-3 and stellar abundances from MWM DR19.
We note that none of the stars in our final sample carry the \texttt{flag\_bad} mask. The stars in our final sample span effective temperatures of 3752–5115 K, surface gravities of 1.31–3.35 dex, and a median signal-to-noise ratio of 184.26. These ranges reflect the expected properties of red giant stars observed with high-quality APOGEE spectra. 
We plot these stars in [Mg/H] vs. [Fe/Mg], colored by asteroseismic age in the left panel of Figure \ref{fig:mg_vs_fe_log_age}.

\section{Predicting Ages from Abundances} \label{sec:abund_age}

Figure~\ref{fig:mg_vs_fe_log_age} shows a strong asteroseismic age gradient in the [Mg/H] vs. [Fe/Mg] plane.
This indicates that stellar abundances can be used as an age indicator.
We observe that older stars tend to have lower [Fe/Mg] ratios (older stars are $\alpha$-enhanced) compared to younger stars, and that older stars tend to have lower values of [Mg/H] (older stars have lower metallicities).
While the trend in Figure~\ref{fig:mg_vs_fe_log_age} is strong, it also shows scattered asteroseismic age outliers across the abundance plane---these are the stars we identify and analyze in this work.

To make outlier identification precise, we construct a simple regression model that predicts asteroseismic ages from surface abundances.
We use a k-nearest neighbors (kNN) regression method with $k=11$.
We build a KD-tree \citep{KNN} using the scipy \citep{2020SciPy-NMeth} implementation with a Euclidean distance metric and a leaf size of two. 
For each star in the dataset, the algorithm calculates the distances between stars in [Mg/H], [Fe/Mg], and $\logg$ and identifies each star's $k=11$ nearest neighbors in this three-dimentional space.
In detail we divide $\logg$ by 10 before computing the neighbor distances with the identity Euclidean metric.
We take the predicted age for each star to be the median asteroseismic age of the $k=11$ neighbors.
We chose the $\logg$ factor and $k=11$ somewhat intuitively; in principle these choices could be carefully optimized.
We refer to this kNN-predicted age as the ``abundance age'' in what follows.
We include $\logg$ in the KD-tree to ensure that neighbors in abundance space are also of similar evolutionary states, improving the accuracy of our age predictions and reducing systematic abundance differences between stars and their neighbors \citep[e.g.,][]{griffith2021}. Including $\teff$ instead of $\logg$ produces a similar distance metric and does not change our results.

\begin{figure}
    \centering
    \includegraphics[width=\linewidth]{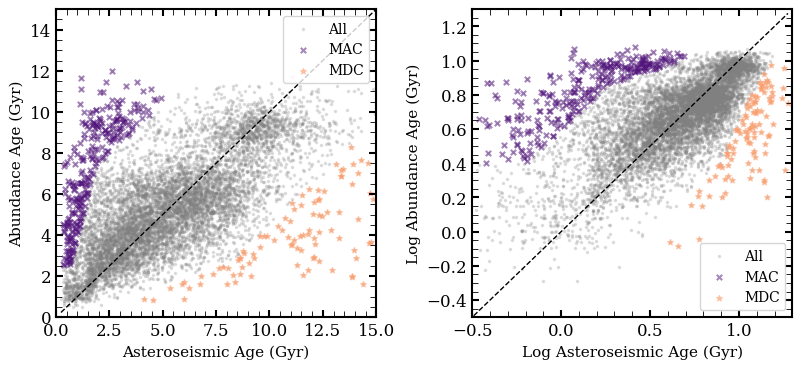}
    \caption{Comparison of asteroseismic ages and abundance ages for all stars in our sample. Left: asteroseismic ages vs. abundance ages in Gyr. Most stars fall near the 1:1 line indicating strong agreement between abundance and asteroseismic ages. Two distinct outlier populations are highlighted: Mass Accretor Candidates (MAC; dark purple crosses), which appear anomalously younger in abundance ages than in asteroseismic ages, and Mass Donor Candidates (MDC; orange stars), which appear older in abundance ages than in asteroseismic ages. The MAC and MDC populations are stars whose surface abundances are inconsistent with their inferred asteroseismic masses, and are selected using both linear and logarithmic residual thresholds (see Section \ref{outlier_section}). Right: Same as left, but with logarithmic ages.}
    \label{fig:cutoffs}
\end{figure}

In the right panel of Figure~\ref{fig:mg_vs_fe_log_age}, we plot the observed stellar [Mg/H] vs. [Fe/Mg] abundances, now colored by the abundance ages. We see the same age trends as observed in the left panel with the asteroseismic ages, but now with far fewer outliers and a smoother age gradient. The abundance ages are robust, with 75\% of stars having abundance ages within 2.2 Gyr of their measured asteroseismic ages. This can be seen in both panels of Figure~\ref{fig:cutoffs}, which compare the asteroseismic and abundances ages on both linear (left) and logarithmic (right) scales. In these figures, we see that most stars fall along the 1:1 relationship, but that large outliers exist among both the asteroseismically young and asteroseismically old stars.

\section{Age-Abundance Outlier Identification} \label{outlier_section}

To identify stars with discrepant asteroseismic and abundance ages, we define two residuals: one in linear space ($\DAge$) and one in logarithmic space ($\DLogAge$) such that
\begin{equation}\label{eq:delta_age}
\Delta \text{Age} = \text{Age}_{\text{asteroseismic}} - \text{Age}_{\text{abundance}}
\end{equation}
and
\begin{equation}\label{eq:delta_log_age}
    \Delta \log_{10}(\text{Age}) = \log_{10}(\text{Age}_{\text{asteroseismic}}) - \log_{10}(\text{Age}_{\text{abundance}}).
\end{equation}

\begin{figure}
    \centering
    \includegraphics[width=\linewidth]{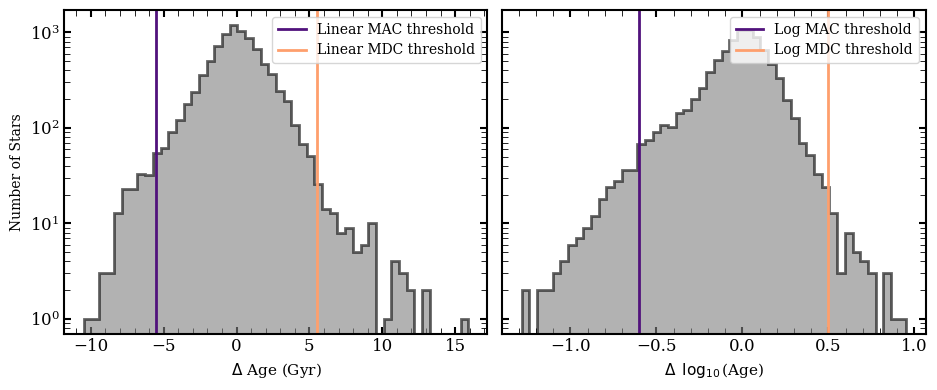}
    \caption{Distributions of residuals between the linear (left) and logarithmic (right) asteroseismic and abundance ages for all stars in our sample. calculated in Equations \ref{eq:delta_age_lin} and \ref{eq:delta_age_log}. We show the respective linear and logarithmic thresholds used to classify MAC (purple) and MDC (orange) outliers.}
    \label{fig:ast-knn}
\end{figure}

These residuals show the discrepancy between the asteroseismic ages and the abundance ages predicted by our kNN regression model (Section~\ref{sec:abund_age}). In Figure~\ref{fig:ast-knn}, we show the distributions of the linear (left panel) and logarithmic (right panel) residuals. We see that while most stars have small residuals ($|\DAge|< \sim 3$ Gyrs), there are stars in the wings of the distributions with $|\DAge| > \sim 5$ Gyrs and $|\DLogAge| > \sim 0.5$.

We consider stars with unusually large residuals as age-abundance outliers. We refer to outliers with a negative residual (i.e., asteroseismic age less than abundance age) as Mass Accretor Candidates (MAC), as these stars may have had their masses inflated and thus their asteroseismic ages deflated by accreting mass in a binary interaction.  Similarly, we refer to outliers with a positive residual (i.e., asteroseismic age greater than abundance age) as Mass Donor Candidates (MDC), as these stars may have had their masses deflated and thus their asteroseismic ages inflated by donating mass in a binary interaction. We classify these outliers using the following thresholds:
\begin{equation}
\label{eq:delta_age_lin}
\Delta \text{Age} =
\begin{cases}
< -5.5 \text{ Gyr} & \text{(MAC)} \\
> +5.5 \text{ Gyr} & \text{(MDC)}
\end{cases}    
\end{equation}
or
\begin{equation}
\label{eq:delta_age_log}
\Delta \log_{10}(\text{Age}) =
\begin{cases}
< -0.6 & \text{(MAC)} \\
> +0.5 & \text{(MDC)}.
\end{cases}
\end{equation}
Linear and logarithmic thresholds are needed, as a purely logarithmic threshold underrepresents stars with large absolute discrepancies at older ages, and a purely linear threshold underrepresents stars with large relative discrepancies at younger ages. These thresholds can be seen in Figure~\ref{fig:ast-knn}. Additionally, we require the predicted or asteroseismic age to be greater than 2.5 Gyrs, as our abundance ages are not precise enough to robustly identify outliers in stars with young asteroseismic and abundances ages.

\begin{figure}
    \centering
    \includegraphics[width=\linewidth]{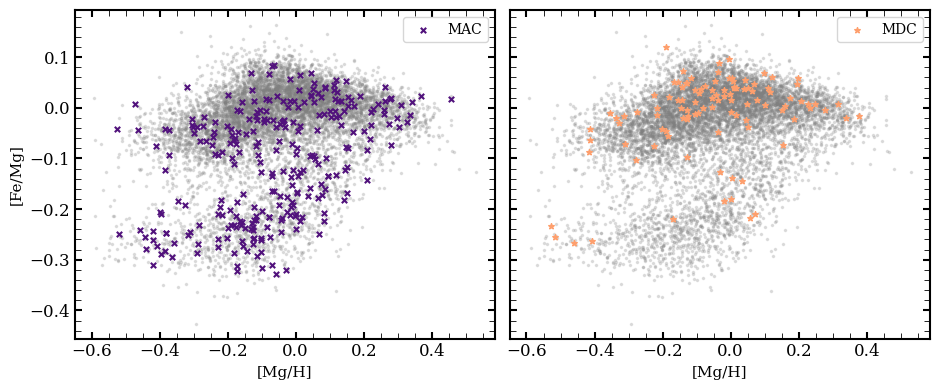}
    \caption{Stellar [Mg/H] vs. [Fe/Mg] for the full stellar sample (grey dots), with MAC (dark purple crosses) and MDC (orange stars) highlighted in the left and right panels, respectively. The MAC and MDC identification is described in Section~\ref{outlier_section}. These age abundance outliers span a broad range of surface abundance values, with MDC concentrated at higher [Fe/Mg] and MAC spanning both low- and high-[Fe/Mg].}
    \label{fig:MAC_MDC_outliers}
\end{figure}

With these requirements we identify 377 total outliers, of which 272 are MAC and 105 are MDC. The MAC and MDC stars represent less than 5\% of our stellar sample. We highlight these stars' asteroseismic vs. abundance ages in Figure~\ref{fig:cutoffs} and show their location in [Mg/H] vs. [Fe/Mg] space in Figure~\ref{fig:MAC_MDC_outliers}. We find that the MDC are concentrated in the high-[Fe/Mg] (low-$\alpha$) population, while the MAC span a wider range of [Fe/Mg] values. The subset of our MAC located in the low-[Fe/Mg] (high-$\alpha$) population are historically referred to as YAR stars. While all of our MAC show younger asteroseismic ages than abundance ages, not all are $\alpha$-rich, prompting a need for our new naming mnemonic. 
Regardless, all outliers identified here have abundance ages that do not align with their asteroseismic ages, suggesting unique stellar histories.

In Table~\ref{tab:outliers} we provide the first lines of our outlier catalog, listing the stars' IDs, abundance ages, and MAC/MDC designation. The full table is available on the online journal.

\begin{table}[h!]
\centering
\begin{tabular}{cccc}
\hline
KIC ID & SDSS ID & Abundance Age & Group \\
\hline
1160789 & 66647306 & 8.6951 & MDC \\
1163359 & 66668648 & 9.5694 & MAC \\
1435573 & 66668319 & 5.7297 & MDC \\
\ldots & \ldots & \ldots & \ldots \\
\hline
\end{tabular}
\caption{Outlier stars with KIC ID, SDSS ID, abundance age, and MAC or MDC classification. The full table is available online.}
\label{tab:outliers}
\end{table}

\section{Exploring Diagnostics of Binarity}\label{sec:explor_params}

In the previous section, we identified outlier stars that exhibit discrepancies between their predicted abundance ages and asteroseismic ages. Since asteroseismic age is inferred from stellar mass, the age discrepancy observed in these outliers suggests that their current masses do not reflect their birth masses, potentially due to binary mass transfer or merger events. If these outlier stars are the product of binary interactions, they may show other signatures of past or ongoing binarity such as radial velocity scatter, rapid rotation, anomalous surface abundances, and excess UV flux. 

In the following subsections, we compare the distributions of such stellar parameters for the MAC and MDC samples with two control samples: their abundance neighbors and their age neighbors.
The abundance neighbors control samples consists of the MAC/MDC star's three nearest neighbors in [Mg/H], [Fe/Mg], and $\logg$ as determined by the KD-tree described in Section~\ref{sec:abund_age}. Similarly, the age neighbors control sample consists of the MAC/MDC star's three nearest neighbors in asteroseismic age and $\logg$. To identify the age neighbors we construct a KD-tree similar to that described in Section~\ref{sec:abund_age}, but with neighbor distances computed using asteroseismic ages rather than abundances. We construct the abundance and age neighbor sample for the MAC and MDC stars separately. 

\begin{figure}[htb!]
    \centering
    \includegraphics[width=\linewidth]{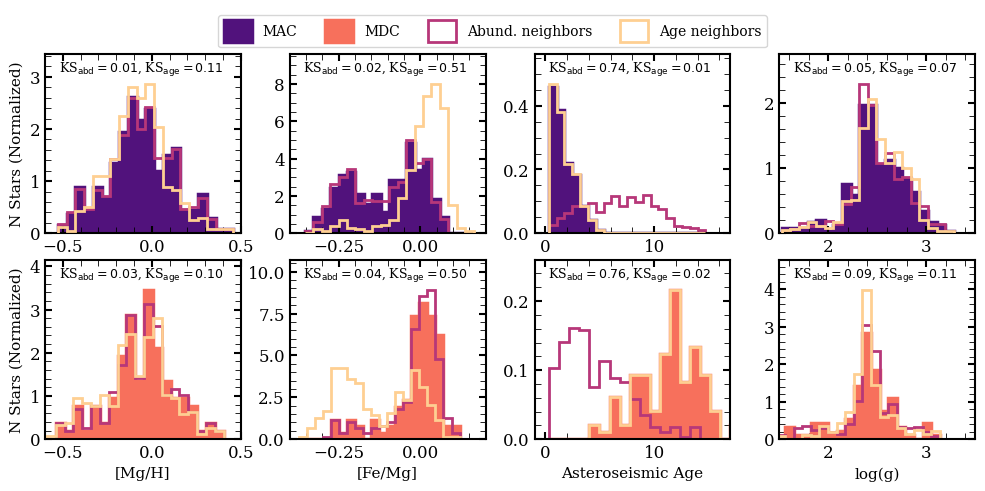}
    \caption{Top row: distributions of stellar parameters for the MAC (dark purple), MAC abundance neighbors (pink) and MAC age neighbors (yellow). Stellar parameters include [Mg/H] (first column), [Fe/Mg] (second column), asteroseismic age (third column) and $\logg$ (fourth column). In each panel we quote the KS statistics between the MAC parameter distribution and that of the abundance (KS$_{\rm abd}$) and age (KS$_{\rm age}$) neighbors. Bottom row: same as top but for the MDC (orange), MDC abundance neighbors (pink), and MDC age neighbors (yellow).}
    \label{fig:kNN_hists}
\end{figure}

Figure~\ref{fig:kNN_hists} shows the distribution of [Mg/H], [Fe/Mg], asteroseismic age, and $\logg$ for the MAC (top row) and MDC (bottom row) alongside their respective abundance and age neighbor samples. These distributions show how the MAC and MDC ages (abundances) differ from their abundance (age) neighbors. We see that the MAC and MAC abundance neighbors have nearly identical [Mg/H] and [Fe/Mg] distributions. This is also seen for the MDC and MDC abundance neighbors. The similarity in abundance distributions is expected. While the age neighbor samples are similar in [Mg/H] for both the MAC and MDC, they differ significantly in [Fe/Mg]. Conversely, the MAC and MAC abundance neighbor age distributions differ, with the MAC having systematically younger ages than their abundance neighbors. The MDC have systematically older ages than their abundance neighbors. Both the MAC and MDC have nearly identical age distributions with their age neighbors, as expected. Finally, we see that both neighbor samples have $\logg$ distributions consistent with the MAC/MDC. 

We quantify the agreement or disagreement between the candidate and neighbor samples with the two-sample Kolmogorov-Smirnov (KS) test.
This is a frequentist test of the hypothesis that two samples are drawn from the same underlying distribution \citep{hodges1958}.
If the samples are drawn from the same underlying distribution, we expect the test to return a small KS statistic, which corresponds to a large $p$-value (greater than $\sim0.01$).
If the samples are not drawn from the same distribution, we expect the test to return a large KS statistic, which corresponds to a small $p$-value.
When the $p$-value is small, that indicates that, in a frequentist sense, it unlikely that the two samples are drawn from the same distribution.
In practice, we report the KS statistic and the logarithm (base ten) of the $p$-value in Table \ref{tab:ks}; $\log p$ values much less than $-2$ might be considered significant, although in what follows we don't impose any precise threshold. 

We quote the KS statistics between the age (KS$_{\rm age}$) and abundance (KS$_{\rm abd}$) neighbor samples for each parameter in the panels of Figure \ref{fig:kNN_hists}.
Confirming the trends we see by eye, we find large KS statistics between the [Fe/Mg] distributions of the candidate and age neighbors (KS$_{\rm age}^{\rm MAC} = 0.51$ and KS$_{\rm age}^{\rm MDC} = 0.50$), as well as the asteroseismic age distributions of the candidate and abundance neighbors (KS$_{\rm abd}^{\rm MAC} = 0.74$ and KS$_{\rm abd}^{\rm MDC} = 0.76$).

\begin{deluxetable*}{ccc|cc|cc|cc}
\tablecaption{Table of KS statistics and log(p) parameters for comparison of  MAC/MDC samples with their abundance and age neighbors.\label{tab:ks}}
\tablehead{
\colhead{} &  \multicolumn{2}{c|}{MAC \& Abd }&\multicolumn{2}{c|}{MAC \& Age }&\multicolumn{2}{c|}{MDC \& Abd }&  \multicolumn{2}{c}{MDC \& Age } \\
\colhead{} &  \multicolumn{1}{c}{KS}&  \multicolumn{1}{c|}{$\log_{10}(p)$}&  \multicolumn{1}{c}{KS}&  \multicolumn{1}{c|}{$\log_{10}(p)$}&  \multicolumn{1}{c}{KS}&  \multicolumn{1}{c|}{$\log_{10}(p)$}&  \multicolumn{1}{c}{KS}& \multicolumn{1}{c}{$\log_{10}(p)$}}
\startdata
 $[\rm Mg/H]$&  0.01&  0.0&  0.10&  -1.6&  0.04&  0.0&  0.10& -0.4\\
 $[\rm Fe/Mg]$&  0.02&  0.0&  0.50&  -47.6&  0.04&  0.0&  0.49& -17.1\\
 Ast. Age&  0.74&  -112.4&  0.01&  0.0&  0.75&  -43.4&  0.02& 0.0\\
 log(g)&  0.05&  -0.2&  0.07&  -0.7&  0.10&  -0.4&  0.11& -0.6\\ 
 v$_\mathrm{scatter}$ & $0.07$ & $-0.3$ & $0.08$ & $-0.4$ & $0.10$ & $-0.1$ & $0.13$ & $-0.3$ \\
RUWE & $0.11$ & $-1.7$ & $0.13$ & $-2.7$ & $0.10$ & $-0.4$ & $0.06$ & $-0.0$ \\
vsin(i)  & $0.29$ & $-0.1$ & $0.28$ & $-0.0$ & $0.33$ & $-0.2$ & $0.14$ & $-0.0$ \\
$\teff$ & $0.27$ & $-12.9$ & $0.08$ & $-0.8$ & $0.14$ & $-1.0$ & $0.12$ & $-0.8$ \\
$[\rm C/N]$ & $0.47$ & $-39.2$ & $0.15$ & $-3.9$ & $0.50$ & $-15.7$ & $0.32$ & $-6.0$ \\
$[\rm O/Mg]$ & $0.15$ & $-3.8$ & $0.17$ & $-4.9$ & $0.21$ & $-2.7$ & $0.19$ & $-2.2$ \\
$[\rm Na/Mg]$ & $0.24$ & $-10.2$ & $0.26$ & $-11.7$ & $0.27$ & $-4.4$ & $0.27$ & $-4.7$ \\
$[\rm Al/Mg]$ & $0.25$ & $-11.1$ & $0.14$ & $-3.0$ & $0.17$ & $-1.6$ & $0.22$ & $-2.8$ \\
$[\rm Si/Mg]$ & $0.06$ & $-0.4$ & $0.47$ & $-40.1$ & $0.09$ & $-0.3$ & $0.44$ & $-12.8$ \\
$[\rm S/Mg]$ & $0.09$ & $-1.3$ & $0.16$ & $-4.5$ & $0.15$ & $-1.2$ & $0.19$ & $-2.0$ \\
$[\rm K/Mg]$ & $0.06$ & $-0.3$ & $0.13$ & $-2.5$ & $0.17$ & $-1.6$ & $0.13$ & $-0.8$ \\
$[\rm Ca/Mg]$ & $0.15$ & $-3.9$ & $0.51$ & $-48.1$ & $0.23$ & $-3.3$ & $0.43$ & $-12.1$ \\
$[\rm Cr/Mg]$ & $0.08$ & $-0.9$ & $0.47$ & $-39.1$ & $0.13$ & $-0.8$ & $0.39$ & $-10.2$ \\
$[\rm Mn/Mg]$ & $0.03$ & $-0.0$ & $0.45$ & $-36.6$ & $0.08$ & $-0.2$ & $0.47$ & $-15.0$ \\
$[\rm Co/Mg]$ & $0.05$ & $-0.1$ & $0.25$ & $-11.0$ & $0.10$ & $-0.3$ & $0.42$ & $-11.3$ \\
$[\rm Ni/Mg]$ & $0.07$ & $-0.5$ & $0.48$ & $-41.7$ & $0.10$ & $-0.4$ & $0.46$ & $-14.3$ \\
$[\rm Ce/Mg]$ & $0.13$ & $-2.8$ & $0.36$ & $-22.5$ & $0.16$ & $-1.4$ & $0.37$ & $-8.9$ \\
 NUV $-$ BP & $0.15$ & $-3.5$ & $0.05$ & $-0.2$ & $0.05$ & $-0.0$ & $0.11$ & $-0.6$ \\
$\Rguide$ (kpc) & $0.05$ & $-0.2$ & $0.25$ & $-11.3$ & $0.17$ & $-1.6$ & $0.28$ & $-5.0$ \\
$\Zmax$ (kpc) & $0.08$ & $-0.8$ & $0.21$ & $-7.8$ & $0.20$ & $-2.6$ & $0.16$ & $-1.4$ \\
\enddata
\tablecomments{Large KS-values indicate that the distributions of a given parameter are unlikely to be drawn from the same distribution, and and small $p$-values ($\logp<-2$) indicates that the difference is significant. }
\end{deluxetable*}

\begin{figure}
    \centering
    \includegraphics[width=1\linewidth]{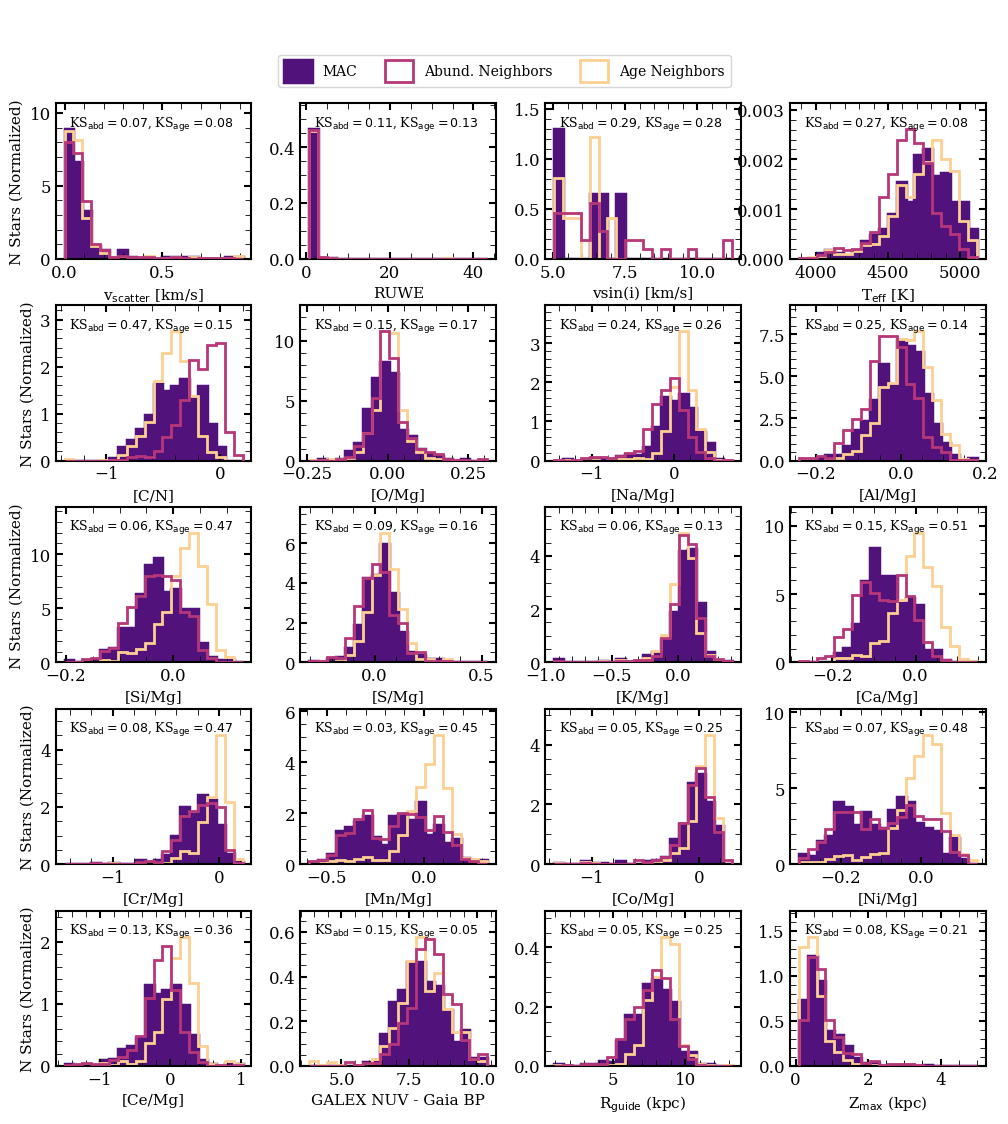}
    \caption{Same as Figure~\ref{fig:kNN_hists}, with the MAC in purple, MAC abundance neighbors in pink, and MAC age neighbors in yellow. Each panel shows the distribution of a different stellar parameter, noted on the $x$-axis.}
    \label{fig:mac_subplot}
\end{figure}

\begin{figure}
    \centering
    \includegraphics[width=1\linewidth]{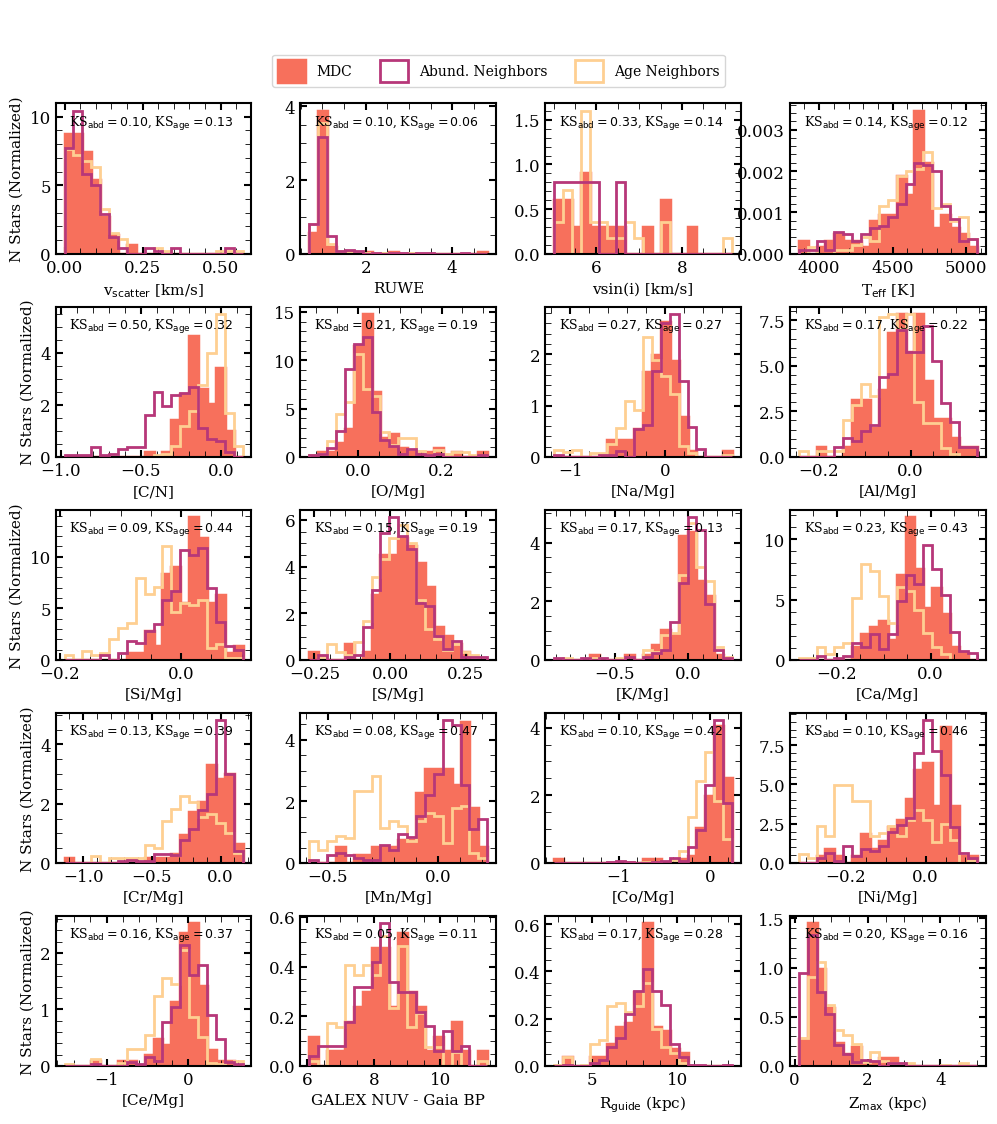}
    \caption{Same as Figure~\ref{fig:mac_subplot}, but for the MDC (orange) and nearest neighbor samples}
    \label{fig:mdc_subplot}
\end{figure}

We conduct a similar comparison of the candidate and neighbor distributions for many stellar parameters. We present the distributions, KS statistics, and $\log_{10}(p)$ values in Figures~\ref{fig:mac_subplot} (MAC) and~\ref{fig:mdc_subplot} (MDC) and Table \ref{tab:ks}. We discuss each stellar parameter in detail below.

\subsection{Radial Velocity and Astrometry}

In a binary stellar system, both stars orbit around a common center of mass. If a portion of this motion is in parallel with the system's radial motion towards or away from us, we will observe periodic red and blue shifts in the spectral features. This will cause variations in the measured RV, with the values dependent upon the configuration of the system. Spectroscopic binaries can thus be identified though their large radial velocity (RV) scatter. This binary diagnostic often requires many spectra taken over a long time baselines. 

In the MWM catalog, many stars have multiple visits and thus multiple RV measurements \citep{meszaros2025, sdssdr19}. If the standard deviation (STD) of the visit RVs  significantly exceeds the typical RV error (\texttt{std\_v\_rad} $>$ 1 km/s and \texttt{std\_v\_rad} $> 5\,\times$ \texttt{median\_e\_v\_rad}\footnote{See \url{https://www.sdss4.org/dr16/irspec/use-radial-velocities/} for binary detection criteria from SDSS-IV.}), there is likely a binary companion present. However, this method is most reliable for stars with more than five high-quality visits. In our sample, 2,222 stars (25\%) have more than 2 visits while only 219 stars (2.5\%) have more than five visits. No stars in our sample have RV scatter greater than 1 km/s, so no stars meet the SDSS criteria for a potential spectroscopic binary.   

Though the MWM RV scatter may not be a robust binary diagnostic for our stars, we examine the parameter distributions for our candidate and neighbor samples. In Figure~\ref{fig:mac_subplot} we show the distributions of the RV scatter for the MAC and the MAC abundance and age neighbors, finding that most stars in all three distributions have an RV scatter $< 0.2$ km/s. We identify 11 MAC with RV scatter $>0.2$ km/s (4\%). A similar percentage of MAC abundance and age neighbors have large scatter as well. The small KS statistics (KS$_{\rm abd}^{\rm MAC} = 0.07$ and KS$_{\rm age}^{\rm MAC} = 0.08$), indicate that the MAC and neighbor samples have consistent RV scatter distributions. Similarly, the MDC and MDC abundance and age neighbors, shown in Figure~\ref{fig:mdc_subplot}, have similar RV scatter distributions and a small KS statistics (KS$_{\rm abd}^{\rm MDC} = 0.10$ and KS$_{\rm age}^{\rm MDC} = 0.13$), with less than 3\% of MDC and MDC neighbor stars having RV scatter $>0.2$ km/s. 

As all stars in our sample have been observed by Gaia, we also check the Gaia DR3 catalog \citep{gaia2016, gaia2023} for signs of binarity. As Gaia is an astrometric mission, measuring the parallaxes and proper motions of billions of stars, it has enabled the detection of millions of binaries \citep[e.g.,][]{elbadry2021}. The renormalized unit weight error (RUWE), a measure of the goodness-of-fit of the Gaia astrometric solution, tends to be large ($\gtrsim 1.4$) for wide binaries \citep[e.g.,][]{belokurov2020, patton2024}.  We compare the candidate and neighbor distributions for the RUWE, shown in Figures~\ref{fig:mac_subplot} (MAC) and~\ref{fig:mdc_subplot} (MDC). As with the MWM RV scatter, we do not see significant differences between the RUWE distributions for the MAC and MAC neighbors nor for the MDC and MDC neighbors, though there are a few stars ($3-10\%$) with large RUWE in each candidate and neighbor sample. 

Finally, we check the Gaia DR3 \texttt{non\_single\_star} flag, which is set for suspect astrometric, spectroscpic, and eclipsing binaries. We find that $11\%$ of MAC are suspect binaries in Gaia, while $<7\%$ of the MAC age and abundance neighbors are flagged for binarity. The \texttt{non\_single\_star} flag is set for $<5\%$ of MDC and MDC neighbors. While this confirms that there is a higher rate of known binary among the MAC than MAC neighbors, it also shows that there is some contamination from binary systems in our abundance and age neighbor samples.

\subsection{Rotational Velocity}

Typical red giants rotate very slowly, with a $\vsini<1$ km/s \citep{medeiros1996}. Binary interactions, however, can spin up stars and cause rotational enhancements in a subset of the population \citep{carlberg2011}. This can occur through main sequence or subgiant mergers, resulting in a rapidly rotating single giant star \citep{peterson1984, leiner2017}, or through tidal synchronization with a binary companion, causing rotational enhancement within a binary system \citep{zahn1989}. Previous studies have identified populations of rotationally enhanced stars in the Kepler field, adopting $\vsini > 5$ km/s as a rapid rotation threshold \citep{tayar2015, patton2024}. 

Rotational velocities can be derived from spectral fits since rotation broadens the absorption features in stellar spectra. However, it is difficult to derive values of $\vsini$ for giant stars, as macroturbulence typically dominates the line broadening. In the ASPCAP spectral reduction pipeline, $\vsini$ is only fit to stars on the main sequence. Adding an additional dimension to the giant star spectral grids leads to poor results and long computation time.  While rotational velocities are not reported in the APOGEE or MWM catalogs, \citet{patton2024} derive values of $\vsini$ for 15,220 giants in the APOKASC sample, following the methods of \citet{tayar2015}. They identify 746 enhanced rotators and report rotational velocities for stars with $\vsini > 5$ km/s---their detection threshold.

Within our sample, 5 MAC and 16 MDC are rotationally enhanced, representing 2\% of the MAC sample and 15\% of the MDC sample. We find that less than 10\% of the MAC and MDC neighbor samples are also rotationally enhanced, indicating that more MDC are rapid rotators than expected from their ages or abundances. We show the distributions of $\vsini$ for stars with $\vsini>5$ km/s in Figures~\ref{fig:mac_subplot} (MAC) and~\ref{fig:mdc_subplot} (MDC). Note that only rotationally enhanced stars are shown in these pannels. 
All other stars have normal rotation rates of $\vsini < 5$ km/s, but the exact values cannot be reliably derived. For the MAC sample, we find $\KSabdMAC=0.29$ and $\KSageMAC=0.28$. Similarly, for the MDC, we find a $\KSabdMDC=0.33$ and $\KSageMDC0.14$. While the KS statsitics for all tests are large, the $\log(p)$ values are near zero and indicate that the we cannot confidently conclude that the distributions are drawn from different parent samples.

\subsection{Stellar Surface Abundances}\label{subsec:abund}

Anomalous stellar abundances, including C and N, can be signs of mass transfer or stellar mergers. While most stellar abundances remain constant through out a star's life, the C and N abundances change as the CNO cycle occurs in the core and dredge up brings CNO processed material to the surface. The photospheric abundances of C and N in giant stars thus changes over the star's lifetime by an amount dependent upon the stellar mass \citep{iben1967}, with dredge up increasing the surface abundance of N and decreasing the surface abundance of C in higher mass stars \cite[e.g.,][]{karakas2014}. Because of this, [C/N] ratios are often used to derive stellar masses and ages. If stars do not follow the typical [C/N] age relation, their surface [C/N] abundance may have been altered by mass transfer or a merger event \citep[e.g.,][]{bufanda2023}. In this section we analyze the MWM abundances for our stellar sample. We exclude Ti, P, and V due to the lower quality of their abundance measurements. We show only unflagged abundance values---removing at most $4\%$ of our sample (for [C/N]).

\begin{figure}
    \centering
    \includegraphics[width=\linewidth]{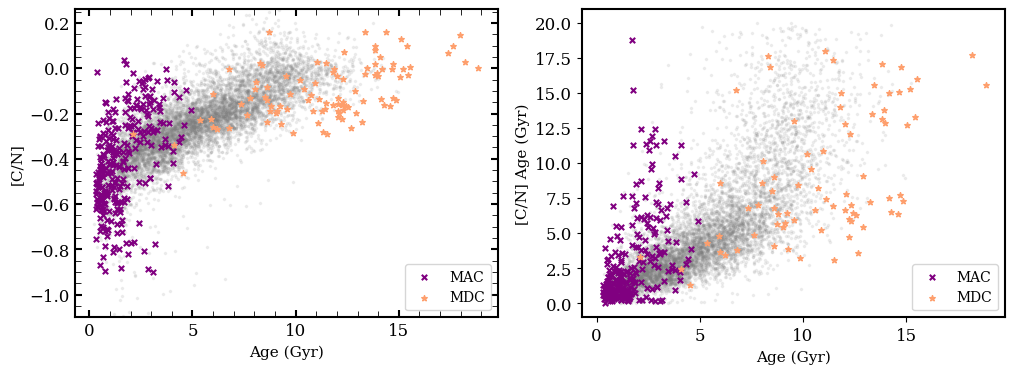}
    \caption{Asteroseismic age vs. [C/N] (on left) and asteroseismic age vs. [C/N] ages (right) for the full sample (grey), MAC (dark purple crosses) and MDC (orange stars).}
    \label{fig:CN_age}
\end{figure}

In Figure~\ref{fig:CN_age}, we show the MWM [C/N] vs. asteroseismic age relationship and the [C/N] age vs. asteroseismic age relationship, highlighting the MAC and MDC. We see the expected trend of [C/N] increasing with age, and plateauing at large ages \citep{roberts2025}. While the MAC and MDC are aligned with this general trend, their distributions appear offset from the main APOKASC sample, with the MAC having higher [C/N], and the MDC having lower [C/N], than stars of similar asteroseismic age. This is seen clearly in Figures~\ref{fig:mac_subplot} and~\ref{fig:mdc_subplot}, where we show the [C/N] distribution for the MAC and MDC, respectively, alongside their abundance and age neighbor samples. 
We calculate [C/N] ages with equations from the APOGEE DR17 abundance calibration \citep{2022AJ....163..229S}, which derived a new empirical relationship between open cluster ages and [C/N] ratios for evolved stars. Here we also see a spread in [C/N] ages for the MAC and MDC that do not align with stars of similar asteroseismic age. We note that the [C/N] tends to over predict the asteroseismic ages of the older stars. This is seen in \citep{2022AJ....163..229S}, but may be exaggerated here as we use DR19 stellar abundances with the DR17 abundance calibration. 

For the MAC, we observe that the [C/N] distribution spans a wide range of values from $-0.9$ to $0.1$, with a median value of [C/N] $\approx-0.4$. This is notably offset from their abundance neighbors (centered at [C/N] $\approx0.0$) and near the center of their age neighbor distribution.
The KS statistics between the MAC and abundance neighbor sample is large with $\KSabdMAC=0.47$. The $\logp$ value significantly less than $-2$ (see Table~\ref{tab:ks}) indicate that it is unlikely that the MAC and abundance neighbor [C/N] distributions are drawn from the same sample. The age neighbor [C/N] distribution is much more similar to the MAC, with $\KSageMAC=0.15$, though the MAC distribution is visibly wider than that of its age neighbors. The range of [C/N] values in the MAC sample suggests a diversity of formation scenarios, as discussed in \citet{2019Hekker} and \citet{2021ApJ...922..145Z} for YAR stars. MAC with low [C/N] may be truly young, or may have merged on main sequence before dredge up began. A main sequence merger would cause an initially less massive star to undergo dredge up as a more massive star, producing surface [C/N] abundances that align with the inflated mass. Conversely, MAC with higher [C/N] may have undergone a mass transfer or merger event during or after dredge up---inflating their mass and deflating their age while retaining the [C/N] surface abundances of a lower mass star. This later scenario is seen for many YAR stars studied by \citet{2023A&A...671A..21J}. 
Additionally, Sinha et al. (in prep) find that close binarity can increase [C/N] by around 0.4 dex, corresponding to an age deflation of about 1 Gyr.

The MDC sample spans a smaller range of [C/N], from roughly $-0.4$ to $0.1$. This distribution is centered on a higher value of [C/N] ($\approx -0.2$) than MDC abundance neighbors and a lower value of [C/N] than the age neighbors. The KS statistics show that the MAC [C/N] distribution is not consistent with either neighbor sample, as $\KSabdMDC=0.50$ and $\KSageMDC=0.32$. The MDC with higher [C/N] may be truly old, low mass stars with anomalously high [Fe/Mg] abundances.  MDC with lower [C/N] abundances, expected of more massive stars, may have had mass stripped after or during dredge up. 

Among the other elements measured by MWM, we find that the [X/Mg] abundance distributions for both the MAC and MDC agree well with their abundance neighbors. This is not surprising, as stars in binary systems tend to have very similar surface abundances \citep{hawkins2020}, so mass transfer may not dramatically alter a star's observed chemical composition for many elements. We include the [X/Mg] distributions for the age neighbor samples Figure~\ref{fig:mac_subplot} (MAC) and ~\ref{fig:mdc_subplot} (MDC), but do not discuss them in detail. We find that the MAC and MAC abundance neighbor distributions have KS statistics of less than 0.2 for [O/Mg], [Si/Mg], [S/Mg], [K/Mg], [Ca/Mg], [Cr/Mg], [Co/Mg], [Ni/Mg], and [Ce/Mg], suggesting that the distributions could be drawn from the same sample. Only [Al/Mg] and [Na/Mg] have a larger KS statistics of 0.24 and 0.25, respectively, with the MAC sample enhanced in [Al/Mg] and [Na/Mg] relative to their abundance neighbors. 

We further find that the MDC and MDC abundance neighbor distributions have KS statistics less than 0.2 for [Al/Mg], [Si/Mg], [S/Mg], [K/Mg], [Cr/Mg], [Mn/Mg], [Co/Mg], [Ni/Mg], and [Ce/Mg]. Among the remaining elements, the MDC [O/Mg] is slightly enhanced relative to the MDC abundance neighbors, but the difference is not very significant as indicated by the $\logp=-2.7$. The MDC [Na/Mg] and [Ca/Mg] are slightly depleted relative to their abundance neighbors, with the [Na/Mg] distribution differences being the most statistically significant. Overall, most [X/Mg] abundance distributions are similar to their abundance neighbor sample, supporting evidence that the MAC and MDC are outliers in asteroseismic age, rather than abundance.

\subsection{UV Excess}

In close binary systems, stars will undergo common envelope evolution as the more massive star evolves from the main sequence (MS) to the giant branch. When the common envelop stage has passed, the system will eventually become a post-common envelope binary, consisting of the WD core of the previously giant star and its companion \citep[e.g.,][]{webbink2008}. The hot, compact remnant will emit more blue light than is expected from a single MS star. Past studies have identified WD binaries through the systems' UV excess \citep[e.g.,][]{parsons2016, hernandez2021}. While most systems contain a lower mass secondary star, recent work by \citet{anguiano2022} has identified systems with UV excess across the HR diagram---showing that post-MS stars with hot companions are also identifiable in this parameter space.

To investigate potential UV excess in our mass transfer candidates, we cross-match our sample with UV sources observed by the Galaxy Evolution Explorer (GALEX). GALEX was a UV space observatory operated by NASA from 2003 to 2011 \citep{martin2005, bianchi2014}, where it conducted All-Sky, Medium, and Deep Imaging Surveys in the near-UV (NUV) and far-UV (FUV). Following its main mission, GALEX was operated privately, and began the Complete All-Sky UV Survey Extension (CAUSE) until 2013. While the FUV detector stopped working in 2009, CAUSE extended NUV observations to the Galactic plane, including the Kepler field. Here, we adopt unflagged NUV magnitudes from the GALEX-CAUSE-Kepler (GCK) survey reported in \citet{olmedo2015}. This catalog contains 668,928 NUV sources, including 475,164 Kepler sources. 

We identify 5,088 stars in our sample with NUV magnitudes in the GCK, 58\% of our data set. We stress that this subset of stars had detectable NUV fluxes, so our analysis is inherently biased against stars with weak NUV emission. In this subsample, we look for NUV excess in the NUV $-$ Gaia BP color, expecting stars with a hot companion to have smaller (bluer) colors. While we anticipate that some MAC may show evidence of a hot companion, it is not obvious that the MDC should show UV excess unless they are actively donating mass to a companion star.  In Figures~\ref{fig:mac_subplot} and~\ref{fig:mdc_subplot} we show the distribution of the NUV $-$ BP color for the MAC and MDC, respectively, compared to their abundnace and age neighbor samples. 
We find both the MAC and MDC color distributions are consistent with their abundance and age neighbors, statistically evident by their smaller KS statistics ($<0.2$). 

\begin{figure}
    \centering
    \includegraphics[width=0.6\linewidth]{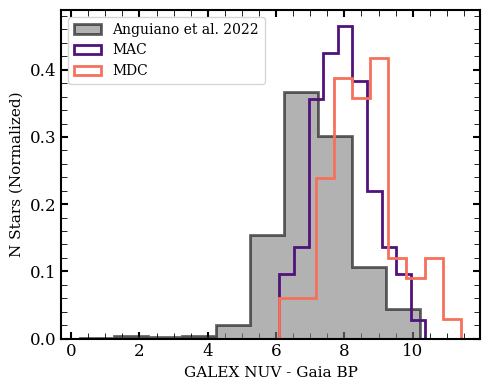}
    \caption{Distribution of GALEX NUV - Gaia BP color for our sample the MAC (dark purple) and MDC (orange) compared with the evolved WD binary candidates from \citet{anguiano2022} (grey, solid).}
    \label{fig:galex_wd}
\end{figure}

In Figure~\ref{fig:galex_wd}, we compare the NUV $-$ BP color distribution of our MAC and MDC sample to the candidate WD binaries identified in \citet{anguiano2022}. This sample consists of 3,414 candidate WD binary systems identified with a combination of GALEX, Gaia, and APOGEE data. We select a subset of 785 targets with an evolved ($\logg<3.5$) companion. Notably, these stars were identified as candidate WD binary systems as they exhibit UV excess in the FUV $-$ NUV color.\footnote{All targets were observed prior to the FUV detector failure in 2009. No stars from the Kepler field are within the \citet{anguiano2022} sample.} In Figure~\ref{fig:galex_wd} we see that the WD binary candidate sample has bluer NUV $-$ BP colors than the MAC or MDC sample, but that the distributions are overlapping. 

The consistency in the NUV $-$ BP color between age-abundance outliers and their neighbors does not support other evidence that many MAC and MDC are mass transfer products. If the MAC and MDC have undergone binary interactions, this result is in tension with \citet{dixon2020}, who identify a statistical difference when comparing the distribution of UV excesses between a sample binary stars and field stars. Alternatively, our results could indicate that WD binaries are difficult to distinguish from field stars in the NUV $-$ BP plane, and that FUV fluxes are needed to more reliably identify WD binary candidates in the \textit{Kepler} field.

\subsection{Orbital Invarients}

Finally, we compare the orbital properties of the mass transfer candidates and their neighbors. While not a direct binary diagnostic, this comparison will determine if the MAC and MDC stellar orbits are most similar to their abundance or age neighbors. In Figures~\ref{fig:mac_subplot} (MAC) and~\ref{fig:mdc_subplot} (MDC) we plot the sample $\Rguide$ and $\Zmax$ distributions, where $\Rguide$ is the orbital radius for a circular, in-plane orbit with the angular velocity of the star and $\Zmax$ is the maximum height above the disk that the star will rise while on its current orbit. These quantities are orbital invariants, similar to actions, but with more intuitive explanations. The orbital parameters are computed in \texttt{gala} \citep{price-whelan2017} using the ``O2GF'' method \citep{sanders2014, sanders2016}, \texttt{MilkyWayPotential2022} mass model, STARHORSE stellar distances \citep{queiroz2023}, and Gaia DR3 astrometry \citep{gaia2023}, as described in \citet{griffith2025}. 

We find that the MAC span a wide range of $\Rguide$, from 5 kpc to 10 kpc with a peak in the distribution near 7.5 kpc. The $\Zmax$ distribution extends to $\Zmax \sim 2$ kpc and peaks near 0.5 kpc, resembling the thick disk. The distributions of the MAC orbital parameters closely resemble those of their abundance neighbors, with $\KSabdMAC=0.05$ for the $\Rguide$ distribution and $\KSabdMAC=0.08$ for the $\Zmax$ distribution. Contrastingly, with KS statistics of 0.25 and 0.21 for $\Rguide$ and $\Zmax$, respectively, the MAC and age neighbor parameters differ and the samples were likely not drawn from the same distribution. We find orbits of the MAC age neighbors occupy larger $\Rguide$ and smaller $\Zmax$, more resembling the thin disk. 

The MDC orbits are most concentrated at $6 \lesssim \Rguide \lesssim 10$ kpc and $\Zmax < 2$ kpc. We find that the orbital parameter distributions of the MDC and their abundance neighbors are similar but not identical, with KS statistics near $0.2$. In $\Rguide$, the KS statistic is larger between the MDC and the age neighbors ($\KSageMDC=0.28$) while in $\Zmax$ the KS statistic is largest between the MDC and abundance neighbors. For this sample, the orbital information cannot constrain if they MDC most similar to their age or abundance neighbors. 

\section{Discussion}\label{sec:discussion}

\begin{figure}
    \centering
    \includegraphics[width=.85\linewidth]{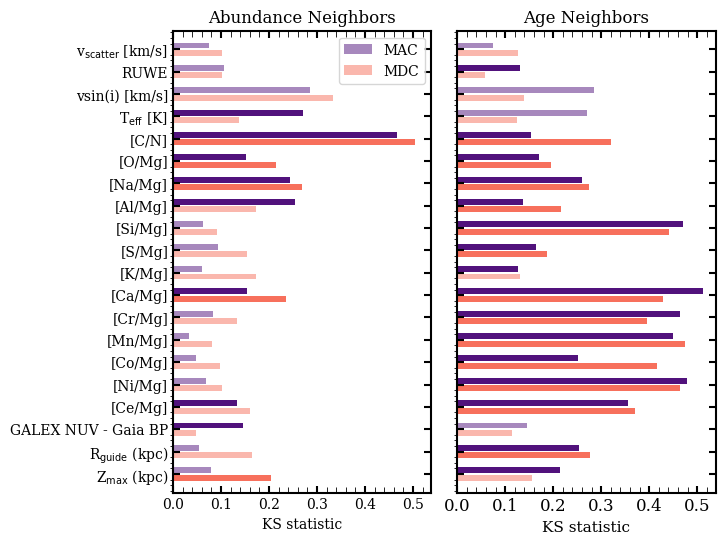}
    \caption{KS statistics between the MAC/MDC (purple/orange) and their abundance (left) and age (right) neighbors. Larger KS statistics indicate more discrepant distributions. Parameters with more significant KS statistics ($\logp<-2$) are shown in the the darker colors while less significant statistics are shown in lighter colors.}
    \label{fig:ks_summary}
\end{figure}

Our analysis shows that while abundance patterns can reliably predict stellar ages for most of the stars in our sample, a subset of stars significantly deviate from the expected trend between age and abundance. In the previous subsections, we explore stellar parameters that may be diagnostic of ongoing or past binarity in these asteroseismic age-abundance outliers. We compare parameter distributions between our MAC (asteroseismic age less than abundance age) and MDC (asteroseismic age greater than abundance age) samples with those of their nearest age neighbors and abundance neighbors, separately. By performing two-population KS tests, we quantify the similarly or dissimilarity of the candidate and neighbor parameter distributions. In Figure~\ref{fig:ks_summary} we show all parameter comparisons presented in this paper, plotting the KS statistics for the MAC/MDC and their abundance neighbors (left panel) and age neighbors (right panel). To convey the significance of the KS statistic, we shade the bars darker for KS tests with $\logp<-2$.

The parameters investigated here can roughly be divided into two categories: those that are direct diagnostics of binarity and should differ between single and binary stars (e.g. RV scatter, rotation rate, UV excess) and parameters that reveal if the candidates are outliers in age or in abundance but are less directly indicators of ongoing or past mass transfer (e.g., stellar surface abundances, orbital invariants). Among the more direct binary diagnostics, we do not see consistent differences between the MAC or MDC (proposed binaries) and the age or abundance neighbor samples (proposed singles). While there are some candidate stars with large RV scatter, rapid rotation, and/or non-single star flags in Gaia, the candidate and neighbor samples are statistically consistent with being drawn from the same parent distributions. This may indicate that the available observations are not able to directly constrain their mass transfer or merger history.

For the indirect indicators of binarity, we generally find that both the MAC and MDC are more similar to their abundance neighbors than their age neighbors, as the KS statistics are smaller and less significant for the candidate-abundance neighbor distributions than the candidate-age neighbors distributions. 
This is in agreement with past studies of YAR stars such as \citet{sun2020}, \citet{2021ApJ...922..145Z}, and \citet{2023A&A...671A..21J}, who generally find that the abundances and orbits of the YAR stars align with those of the thick disk. 
The similarity of these parameters between the candidate and abundance neighbors suggests that the MAC and MDC are outliers in asteroseismic age, rather than abundance. Our work supports supports the existing evidence that many MAC and MDC stars are products of binary mass transfer, inflating/deflating their asteroseismic age while leaving most of their surface abundances unchanged. 

Notably, [C/N] and [Na/Mg] are very significant outliers (KS $>0.2$ and $\logp < -4$) to this trend for both the MAC and MDC, and [Al/Mg] is an outlier for the MAC only. As discussed in Section~\ref{subsec:abund}, [C/N] is an indicator of age, with the surface [C/N] ratio changing as a star goes through dredge up. In YAR stars, \citet{2016A&A...595A..60J} find that [C/N] ratios don't align with single star evolution isochrones, likely due to mass transfer before the donor star's first dredge up process that leaves the accreting star with lower overall [C/N] than an older sample. \citet{2023A&A...671A..21J} find that the two undermassive stars in their sample (similar to our MDC) have high [C/N] ratios compared to the high-$\alpha$ population. We observe that many of our MAC, which have young asteroseismic ages, have lower [C/N] ratios than their abundance neighbors, as found by \citet{grisoni2024}, while the MDC, which have old asteroseismic ages, tend to have larger [C/N] than their abundance neighbors. This places both populations [C/N] distributions closer to their age neighbors, indicating that mass transfer occurred before the stars underwent dredge up, or that some stars may be truly young/old.

The disparity in the [Na/Mg] and [Al/Mg] candidate and abundance neighbor distributions could also indicate mass transfer. We observe the MAC to be enriched in Na and Al compared to their abundance neighbors, which could occur if Na and Al rich material is accreted during the binary interaction. Empirical studies of stellar abundances have found that Na and Al appear to be produced by AGB stars \citep[e.g.,][Ness in prep.]{griffith2019}, supported by recent studies of AGB yields and GCE models \citep[e.g.,][]{karakas2014,kobayashi2020}. It is not clear why the MDC would be depleted in Na only relative to their abundance neighbors. If a large portion of the population was truly old, we would expect more abundance distributions to be shifted towards those of their age neighbor sample.

Overall, we find broad evidence that many of the MAC and MDC experienced binary mass transfer, with $17\%$ of MAC and $20\%$ of MDC showing evidence of ongoing or past binary interactions in Gaia data or rotational velocities. Notably, some ($10-15\%$) MAC and MDC abundance and age neighbors also show these signs of binarity. While we cannot robustly constrain the binary fraction in our population, the parameter distributions explored in this work suggest that the age-abundance outliers are predominately a result of binary interactions. 

However, the broad [C/N] distribution for the MAC suggests that some of these object may truly be age outliers for their surface abundances. Past works have suggested that the YAR stars may have formed due to fluctuations in the SNIa rate due to radial migration \citep{johnson2021}, recent accretion events \citep{sun2023}, or enrichment from inert gas trapped near the end of the Galactic bar \citep{Chiappini2015}. Observationally, \citet{lu2025} identify ten YAR dwarf stars in the Kepler and K2 fields with no signs of mass transfer or merger history---supporting evidence that some YAR stars are truly young while others have had their masses inflated from binary interactions. 

Conversely, radial velocity monitoring campaigns, such as \citet{2023A&A...671A..21J}, have successfully found binary companions for a handful of YAR stars. We compare our sample to this follow up work, and find that our quality and sample cuts contain 31 of the 41 stars they observe. Of the 17 stars in our sample labeled by \citet{2023A&A...671A..21J} as young $\alpha$-rich, 11 are contained in our MAC sample and an additional four sit just below our outlier threshold. \citet{2023A&A...671A..21J} confirm that five of the stars in their YAR sample and our MAC sample are currently in a binary system, and find no evidence of ongoing binarity in an additional three.

More generally, follow-up observing campaigns are needed to better classify all age-abundance outliers. High-resolution, optical spectra could provide additional neutron capture abundances, testing if MAC are enriched in heavy elements produced by AGB stars. Additional FUV fluxes could be combined with NUV flux from GALEX to more reliably identify stars with a hot companion \citep[e.g.,][]{anguiano2022}. Finally, additional high cadence RV monitoring, as that done by \citet{2016A&A...595A..60J, 2023A&A...671A..21J}, could more robustly identify spectroscopic binarity and quantify the fraction of asteroseismic age-abundance outliers that have undergone past mass transfer.

\section*{Acknowledgments}   

We thank Borja Anguiano for sharing the UV fluxes and WD binary candidates from \citep{anguiano2022} and Adrian Price-Whelan for providing stellar orbital parameters. We thank Marc Pinsonneault, Rachel Patton, Adam Wheeler, Lucy Lu, Alexander Stone-Martinez, Johanna Müller-Horn and the Darling research group at CU Boulder for valuable discussions and help on this project. 

E.J.G. acknowledges support for this work provided by NASA through
the NASA Hubble Fellowship Program grant No. HST-HF2-51576.001-A
awarded by the Space Telescope Science Institute, which is operated
by the Association of Universities for Research in Astronomy, Inc.,
for NASA, under the contract NAS 5-26555. E.J.G. was supported during part of this work by an NSF Astronomy and Astrophysics Postdoctoral Fellowship under award AST-2202135. The Flatiron Institute is a division of the Simons Foundation.

Funding for the Sloan Digital Sky Survey V has been provided by the Alfred P. Sloan Foundation, the Heising-Simons Foundation, the National Science Foundation, and the Participating Institutions. SDSS acknowledges support and resources from the Center for High-Performance Computing at the University of Utah. SDSS telescopes are located at Apache Point Observatory, funded by the Astrophysical Research Consortium and operated by New Mexico State University, and at Las Campanas Observatory, operated by the Carnegie Institution for Science. The SDSS web site is \url{www.sdss.org}.

SDSS is managed by the Astrophysical Research Consortium for the Participating Institutions of the SDSS Collaboration, including the Carnegie Institution for Science, Chilean National Time Allocation Committee (CNTAC) ratified researchers, Caltech, the Gotham Participation Group, Harvard University, Heidelberg University, The Flatiron Institute, The Johns Hopkins University, L'Ecole polytechnique f\'{e}d\'{e}rale de Lausanne (EPFL), Leibniz-Institut f\"{u}r Astrophysik Potsdam (AIP), Max-Planck-Institut f\"{u}r Astronomie (MPIA Heidelberg), Max-Planck-Institut f\"{u}r Extraterrestrische Physik (MPE), Nanjing University, National Astronomical Observatories of China (NAOC), New Mexico State University, The Ohio State University, Pennsylvania State University, Smithsonian Astrophysical Observatory, Space Telescope Science Institute (STScI), the Stellar Astrophysics Participation Group, Universidad Nacional Aut\'{o}noma de M\'{e}xico, University of Arizona, University of Colorado Boulder, University of Illinois at Urbana-Champaign, University of Toronto, University of Utah, University of Virginia, Yale University, and Yunnan University.  

This paper includes data collected by the Kepler mission and obtained from the MAST data archive at the Space Telescope Science Institute (STScI). Funding for the Kepler mission is provided by the NASA Science Mission Directorate. STScI is operated by the Association of Universities for Research in Astronomy, Inc., under NASA contract NAS 5–26555.

This work has made use of data from the European Space Agency (ESA) mission
{\it Gaia} (\url{https://www.cosmos.esa.int/gaia}), processed by the {\it Gaia}
Data Processing and Analysis Consortium (DPAC,
\url{https://www.cosmos.esa.int/web/gaia/dpac/consortium}). Funding for the DPAC
has been provided by national institutions, in particular the institutions
participating in the {\it Gaia} Multilateral Agreement.

\software{
Astropy \citep{astropy2013, astropy2018, astropy2022},
astroquery \citep{astroquery},
gala \citep{Price-Whelan:2017}, 
Matplotlib \citep{hunter2007}, 
NumPy \citep{harris2020}, 
Pandas \citep{pandasa, pandasb}, 
scipy \citep{2020SciPy-NMeth}
}

\bibliography{mt}{}

\bibliographystyle{aasjournalv7}
\end{document}